# One-step atmospheric pressure synthesis of the ground state of Fe based LaFeAsO$_{1-\delta}$ superconductor

V. P.S. Awana[*], Arpita Vajpayee, Monika Mudgel, Anuj Kumar, R.S. Meena, Rahul Tripathi, Shiv Kumar, R.K. Kotnala and Hari Kishan

National Physical Laboratory, Dr K.S. Krishnan Road, New Delhi-110012, India

*e-mail: awana@mail.nplindia.ernet.in; Web page: www.freewebs.com/vpsawana

**Abstract**

We report an easy and versatile one-step route of synthesis for newly discovered Fe based superconductor LaFeAsO$_{1-\delta}$ with $0.0 \leq \delta \leq 0.15$. Instead of widely used high-pressure-high-temperature (*HPHT*) synthesis, we applied the normal atmosphere solid-state reaction route. The stoichiometric mixtures of Fe, La$_2$O$_3$, La and As in ratio LaFeAsO$_{1-\delta}$ with $0.0 \leq \delta \leq 0.15$ are sealed in an evacuated quartz tube and further heated at 500, 850 and 1100 $^0$C in Ar for 12, 12 and 33 hours respectively in a single step. The resulting compounds are single phase LaFeAsO crystallized in tetragonal *P*4/*nmm* structure. These samples showed the ground state spin density wave (*SDW*) like metallic behavior below around 150 K. In conclusion the ground state of newly discovered Fe based superconductor is synthesized via an easy one-step solid-state reaction route.

**Introduction**

The search for new superconducting materials got a boost after the invention of high $T_c$ superconductivity in 1987 by Muller and Bednorz [1]. Soon after various Cu based high $T_c$ superconducting (*HTSc*) compounds were invented with their critical transition temperatures ranging from 20 to 134 K [2-4]. However this search remained strictly confined to cuprates until the invention of superconductivity in MgB$_2$ at 40 K in year 2001 [5]. Later the superconductivity was observed at around 5 K in oxy-cobalt hydrate (Na$_x$CoO:H$_2$O) [6]. Another important compound outside the popular cuprates family was $Sr_2RuO_4$ with triplet pairing [7]. As far as the pairing mechanisms are concerned though the $MgB_2$ still seems to follow the strong electron phonon coupling, the others; in particular the *HTSc* cuprates, are still a scientific mystery for the theoreticians [8]. In this direction very recent reports on superconductivity of up to 55 K in REFeAsO (RE = La, Pr, Sm, Nd, Gd) had renewed the interest of scientific community to search for high Tc superconductors outside the cuprates family [9-24].

The recent compound i.e. REFeAsO is the only known superconductor yet having its $T_c$ outside the so-called strong *BCS* (Bardeen Cooper and Schreifer) limit i.e. 40 K. Further the normal state resistivity behavior and with some other features of the REFeAsO are very similar to that of *HTSc* cuprates [18-21]. The Fe based compound provides an opportunity to the theoreticians to think out side the cuprate families in search for the mechanism of high $T_c$ superconductivity [22-24]. The newly discovered Fe based superconducting material is mainly synthesized by the high pressure high temperature (*HPHT*) process with pressure as high as 6 Gpa at 1150 $^0$C [9-12]. Few scant reports are for the normal pressure synthesis as well, but with complicated two or three step reaction routes [13-17]. In the current short rapid communication, we report an easy and versatile single step route for the synthesis of the ground state of the Fe based superconductor $LaFeAsO_{1-\delta}$ with $0.0 \leq \delta \leq 0.15$.

**Experimental**

Stoichiometric amounts of better than 3 N purity of As, Fe, La metal and $La_2O_3$ were weighed and mixed thoroughly in formula ratio $LaFeAsO_{1-\delta}$ with $0.0 \leq \delta \leq 0.15$. For example in case of $LaAsFeO_{0.9}$ the stoichiometric amounts used are: $Fe+As+0.3La_2O_3+0.4La$. The weighed and mixed powders are sealed in evacuated (better than $10^{-4}$ Torr) quartz tubes. The sealed quartz tubes containing various respective samples are heated at 500, 850 and 1100 $^0$C in Ar for 12, 12 and 33 hours respectively in a single step. The x-ray diffraction patterns of these compounds are taken on Rigaku mini-flex diffrractometer. The resistivity measurements are carried out by four-probe method on a close cycle refrigerator in temperature range of 12 to 300 K.

**Results and Discussion**

Figure 1 depicts the X-ray diffraction (*XRD*) patterns of fitted and observed LaFeAsO$_{0.9}$. The XRD patterns of the compound is fitted on the basis of tetragonal, *P*4/*nmm* space group. Besides the main phase (tetragonal *P*4/*nmm*) some very small intensity un-reacted lines arising from either FeAS, or LaAs are also seen in the *XRD* pattern. Worth mentioning is the fact that quality of our one-step atmospheric pressure synthesized material is as good as the *HPHT* or the complicated multi-step route [9-21,25]. The Lattice parameters are: $a$ = 4.03421(23)Å and $c$ = 8.73545(74)Å for LaFeAsO$_{0.9}$. The co-ordinates positions and the quality of fitting parameters are given in Table 1. The *XRD* fitting of other samples of LaFeAsO$_{1-\delta}$ with $0.0 \leq \delta \leq 0.15$ is same to that as observed in Fig. 1 for LaFeAsO$_{0.9}$. The lattice parameters for all the studied samples are tabulated in Table 2. With increase in oxygen vacancies the $a$ and $c$ lattice parameters and unit cell volume decrease continuously. This is in agreement with a recent report on LaFeAsO$_{1-\delta}$ [25].

The Resistance versus temperature (*R-T*) plot for the LaFeAsO$_{0.85}$ sample is shown in Figure 2. The resistance behaviour is metallic from room temperature down to 250 K and later is semiconductor like till 150 K, below 150 K a shallow metallic step is seen down to 90 K and than again semiconducting down to 12 K. The 150 K shallow metallic step is clear indication of the spin density wave (*SDW*) transition of the system [9,13, 16, 19]. It is known that the ground state of this newly discovered Fe based superconductor is magnetic with *SDW* character [9-19, 22-24]. With induction of electron or hole carriers either by F doping [9-21,24] or aliovalent substitutions [16,25,27], the superconductivity can be introduced with $T_c$ of up to 26 K. For our sample the *SDW* character is reminiscent in conductivity measurements as a metallic shallow step below 150 K. Though for sack of brevity the R-T plot of only LaFeAsO$_{0.85}$ sample is shown in Fig.2, but the *SDW* characteristic metallic step is seen in all LaFeAsO$_{1-\delta}$ samples with $0.0 \leq \delta \leq 0.15$. Further all these samples are crystallised in single phase. Our results demonstrate that the *SDW* ground state of LaFeAsO$_{1-\delta}$ is versatile and stable over a wide range of oxygen content. The method of synthesis applied by us is easy and versatile and could be tailored for F doped superconducting REFeAsO$_{1-\delta}$ (RE = La, Pr, Sm, Nd, Gd) compounds.

In summary, the ground state of newly discovered LaFeAsO superconductor was synthesized via an easy and versatile one step route over a wide range of oxygen content. The method applied by us is unique and versatile and hence can be tailored easily for the F doped or substituted LaFeAsO superconductor.

Table 1. Reitveld refined parameters for LaFeAsO$_{0.90}$.

Space group: *P4/nmm*, Lattice parameters ; *a* = 4.0342 (2)Å, *c* = 8.7354 (7)Å

| Atom | Site | x | y | z |
|------|------|------|------|-----------|
| La | 2c | 0.25 | 0.25 | 0.1398(4) |
| Fe | 2b | 0.75 | 0.25 | 0.5 |
| As | 2c | 0.25 | 0.25 | 0.6508(6) |
| O | 2a | 0.75 | 0.25 | 0 |

*Rp*: 5.54%, *Rwp*: 7.37%, *Rexp*: 3.32%, $\chi^2$: 4.92

Table 2. Lattice parameters and cell volume of LaFeAsO$_{1-\delta}$ ($\delta$=0.0 - 0.15).

| Sample | *a*(Å) | *c*(Å) | V(Å$^3$) |
|--------|-----------|-----------|-------------|
| LaFeAsO | 4.0363(3) | 8.7356(7) | 142.322(11) |
| LaFeAsO$_{0.90}$ | 4.0342(2) | 8.7354(7) | 142.168(17) |
| LaFeAsO$_{0.85}$ | 4.0321(7) | 8.7353(24) | 142.023(54) |

**Acknowledgement**

Authors thank their director Professor Vikram Kumar for encouragement. Professor O.N. Srivastava from Banaras Hindu University (BHU) is acknowledged for various fruitful discussions and encouragement.

**Figure Captions**

Fig. 1: Figure 1: Fitted and observed X-ray diffraction patterns of LaFeAsO$_{0.9}$.

Fig. 2: R(T) of the LaFeAsO$_{0.85}$, the SDW transition at around 150K is marked.

Fig. 1

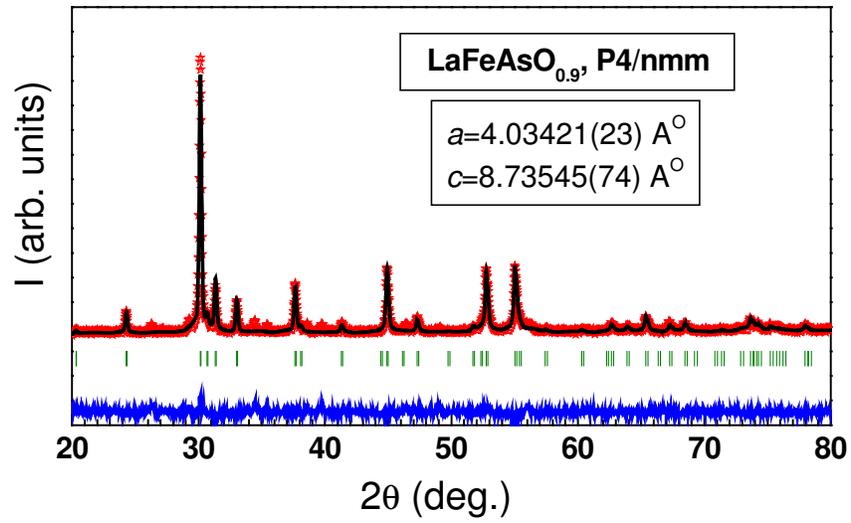

Fig. 2

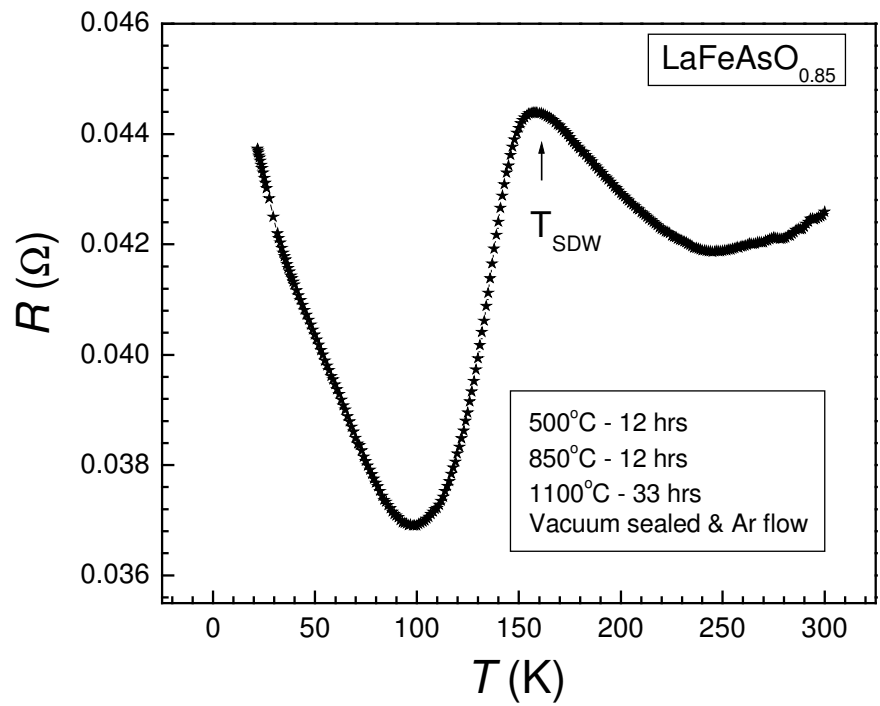